\PassOptionsToPackage{dvipsnames,table}{xcolor}
\documentclass{article}
\usepackage{arxiv}
\usepackage[english]{babel}
\usepackage[utf8]{inputenc}
\usepackage[T1]{fontenc}
\usepackage{xurl}
\usepackage{pdfpages}

\DeclareUnicodeCharacter{2061}{}
\DeclareUnicodeCharacter{2264}{}
\DeclareUnicodeCharacter{03BC}{}

\usepackage{authblk}
\usepackage{makecell}
\usepackage{hyperref}
\hypersetup{
  colorlinks   = true, 
  urlcolor     = black, 
  linkcolor    = blue, 
  citecolor   = black 
}
\usepackage{graphicx} 
\usepackage{amsmath}
\usepackage{mathtools}
\usepackage{xcolor}

\usepackage{color}

\usepackage{booktabs}
\usepackage{makecell}
\usepackage{threeparttable}

\begin{document}
\title{Start from the End: A Framework for Computational Policy Exploration to Inform Effective and Geospatially Consistent Interventions applied to COVID-19 in St. Louis}

\author[1]{David O'Gara \thanks{To whom correspondence should be addressed: david.ogara@wustl.edu}}
\author[2]{Matt Kasman}
\author[3]{Matthew D. Haslam}
\author[1,2,4,5]{Ross A. Hammond}
\affil[1]{Division of Computational and Data Sciences\\Washington University in St. Louis\\ 
St. Louis, MO}
\affil[2]{Center on Social Dynamics and Policy\\
Brookings Institution\\
Washington, DC}
\affil[3]{Department of Health\\City of St. Louis\\St. Louis, MO}
\affil[4]{School of Public Health\\Washington University in St. Louis\\ 
St. Louis, MO}
\affil[5]{Santa Fe Institute\\Santa Fe, NM}
\renewcommand{\thefootnote}{\fnsymbol{footnote}}

\maketitle
\begin{abstract}
    Mathematical models are a powerful tool to study infectious disease dynamics and intervention strategies against them in social systems. However, due to their detailed implementation and steep computational requirements, practitioners and stakeholders are typically only able to explore a small subset of all possible intervention scenarios, a severe limitation when preparing for disease outbreaks. In this work, we propose a parameter exploration framework utilizing emulator models to make uncertainty-aware predictions of high-dimensional parameter spaces and identify large numbers of feasible response strategies. We apply our framework to a case study of a large-scale agent-based disease model of the COVID-19 ``Omicron wave'' in St. Louis, Missouri that took place from December 2021 to February 2022. We identify large numbers of response strategies that would have been estimated to have reduced disease spread by a substantial amount. We also identify policy interventions that would have been able to reduce the geospatial variation in disease spread, which has additional implications for designing thoughtful response strategies.
\end{abstract}

\section{Introduction}
\label{sec:intro}
Epidemics of communicable disease have the capacity to exact a heavy toll on our society. Before, during, and after waves of epidemic spread, policymakers and stakeholders have used mathematical models of disease transmission to forecast potential scenarios, inform mitigation strategies, and explore how we can prepare for the next one \cite{epstein_modelling_2009,howerton_evaluation_2023,loo_us_2024}. An important subset of available models are individual or ``agent-based'' models (ABMs), which explicitly simulate the interactions of individual actors in a population, offering opportunities to understand mechanisms of disease transmission and opportunities for intervention \cite{hammond_overview_2021}. These types of computational models have a long history of informing policy response planning for a multitude of diseases in the United States and worldwide \cite{aylett-bullock_j_2021,eubank_modelling_2004,ferguson_report_2020,ferguson_strategies_2006,germann_mitigation_2006,halloran_modeling_2008,hinch_openabm-covid19agent-based_2021,hladish_projected_2016,kerr_controlling_2021,longini_ira_m_containing_2005,ozik_population_2021,penny_public_2016}.

A significant complication in application of this approach arises when policymakers and stakeholders must balance multiple competing objectives in evaluating potential responses: for example, not just limiting the total outbreak size but also minimizing economic and social disruption and remaining conscious of healthcare system capacity. This type of balancing act presents a steep challenge, which becomes even more difficult as more objectives are considered. Importantly, when assessing interventions in the face of multiple objectives, a globally optimal, one-size fits all solution likely does not exist \cite{binois_portfolio_2025,garnett_bayesian_2023}. Agent-based models have the potential to help policymakers explore what action to take, and understand trade-offs, substitutions, and limitations for any proposed policy interventions.

Reaching this potential requires overcoming technical hurdles. Due to the level of detail embedded in these models, the simulation and experimentation process can be quite expensive, sometimes requiring minutes or hours to complete a single model run \cite{aylett-bullock_j_2021,germann_mitigation_2006}. Therefore, when faced with the question of how to intervene on a complex dynamical system of disease spread while simultaneously satisfying multiple objectives, modeling teams are limited by their computational ``budget,'' or the number of simulations they can conduct in a reasonable amount of time. These efforts are further complicated when epidemic and social dynamics depart from tidy canonical modeling assumptions such as an outbreak stemming from an index case in an otherwise completely susceptible population, complete and durable immunity conferred by vaccination, and perfect adherence to non-pharmaceutical interventions \cite{hunter_taxonomy_2017,shastry_policy_2022}. These assumptions and others did not apply to the late 2021 through early 2022 COVID-19 “Omicron wave” in the United States, resulting in the largest rates of infection over any similar time period \cite{badillo-goicoechea_global_2021,fridman_association_2020,king_time_2021,pro_us_2021,troiano_vaccine_2021,yan_measuring_2021}. Because the Omicron wave happened after the initial stages of the pandemic, differential access to interventions and protective behaviors likely drove disparities in disease spread \cite{faust_racial_2024,kim_social_2020,lundberg_covid-19_2023,mody_quantifying_2022,oates_association_2021}. As a result, when planning for mitigation strategies, the question of what to do, as well as when and how to do it, becomes unclear, and the notion of a one-size fits all solution is likely not applicable in these circumstances. Complex adaptive systems models in general, and ABMs in particular, are well-suited to  modeling scenarios when policy impacts are not uniform, with differential effects across sub-populations or settings \cite{combs_modelling_2019,hammond_development_2020,kasman_childhood_2023,kasman_best_2022,luke_tobacco_2017,kasman_matt_human_2024}. 

Here, we introduce an approach to advance ABM’s capacity to explore the vast policy landscape of potential mitigation strategies, and illustrate the approach with an applied case study of a highly detailed ABM of COVID-19 spread in St. Louis, Missouri, the ``TRACE-STL'' model, which has already informed policy response planning in the greater metro area \cite{hammond_modeling_2021}. Using available data, we parameterize TRACE-STL to look retrospectively at the Omicron wave and explore ten different intervention strategies across vast implementation intensities, allowing us to investigate which mixtures of policies might have been effective in mitigating disease spread while considering multiple (potentially competing) objectives for outcomes. We do so via the use of emulator (or ``surrogate'') models, a class of statistical models used to approximate expensive computer simulations, which can offer uncertainty-aware predictions at unseen model inputs, and guide model exploration across parameter space \cite{garnett_bayesian_2023,gramacy_surrogates_2020}.

A deep and mature literature for calibrating complex simulation models to available data exists, in which emulator models have been used to great effect \cite{fadikar_calibrating_2018,katzfuss_scaled_2022,kennedy_bayesian_2001,ogara_improving_2024,ozik_population_2021,panovska-griffiths_machine_2023,reiker_emulator-based_2021,shattock_impact_2022,vernon_galaxy_2010,vernon_bayesian_2022}. To our knowledge, the use of emulators to identify large regions of candidate policy configurations has not yet been explored. 

Our results indicate that thoughtful emulation is a powerful framework to explore the TRACE-STL model. We identify a large range of policy combinations, many of which are combinations of several lower intensity policies, which would be estimated to have reduced disease spread. We also study our framework’s ability to identify tradeoff and substitution effects for policies across wide ranges. Finally, we also demonstrate that our emulation strategy can not only lead to reductions in total disease spread but can also identify policies that are estimated to reduce the variation in disease burden across census tracts, offering support for the possibility of developing geospatially consistent interventions. Specifically, equalizing disease burden across census tracts or neighborhoods is of high importance to avoid concentrated social or economic costs, to prevent strain on local healthcare systems, and to prevent secondary epidemiological flare-ups. Taken together, our approach may be thought of as a detailed ``response playbook'' that could help policymakers plan and explore myriad epidemic response scenarios.
\section{Methods}
\label{sec:methods}
\subsection{TRACE-STL Model Description}

Our complex simulation setting is an updated version of the ``TRACE-STL'' agent-based model of COVID-19, parameterized and calibrated to the ``Omicron wave'' of disease spread from December 21, 2021 to February 19, 2022. TRACE-STL was originally developed to forecast robust policy response scenarios over a wide range of potential interventions and epidemiological scenarios \cite{hammond_modeling_2021}. A full description may be found in the Supplement, but we recap the main components and model dynamics here. Briefly, TRACE-STL is a large-scale computational disease transmission model of 2.4 million agents in St. Louis, Missouri and its adjoining counties. The synthetic population and contact networks are based on the 2010 RTI Synthetic population data \cite{wheaton2009synthesized,rti_international_2010_nodate}. We use publicly available data on the number of cases, deaths, and tests \cite{noauthor_govexcovid-19_nodate,usa_facts_us_2020,reinhart_open_2021} , mobility trends \cite{google_covid-19_nodate}, vaccinations and boosting \cite{cdc_vaccine_covid-19}, and census-tract SVI \cite{cdc_svi_2024} and simulate a 60-day period where each model timestep denotes one simulated day.

\subsection{Policy Interventions}

We simulate 10 different policy interventions, which are designed to reflect a wide range of potential policy scenarios. Each of the policies are varied over wide ranges, meant to reflect the status quo (at their lowest intensity) and a massive increase (at their highest intensity). Importantly, the policies affect the agent population in different ways: for example, a change in the frequency of diagnostic PCR testing is expected to have a different effect than administering vaccines. We note that we do not simulate social distancing policies, such as school closures and stay-at-home directives. Prior work with TRACE has explored the efficacy of such interventions \cite{ogara_traceomicron_2023}, but under the assumption that these interventions would be infeasible or prohibitively expensive, we did not include them in our experiments. Our policy scenarios and ranges are reported in Table~\ref{tab:policies}. We denote five of our policies as policies regarding changes in ``coverage'' i.e. the amount of an intervention available and applied (such as the number of vaccines) and five of our policies as changes in ``behavior.'' These designations are not meant to be completely descriptive or mutually exclusive but are given for ease of presentation and to demonstrate the many ways policies effect agent dynamics. For example, we note that an individual receiving a vaccine is likely at least a function of both product availability as well as individual behavior.

\begin{table}[]
\begin{tabular}[\textwidth]{lll}
\textbf{Policy}      & \textbf{Description}                & \textbf{Range}             \\
\midrule
\rowcolor[HTML]{F2F2F2} 
\textit{Coverage} &  & \\
PCR tests per day    & Multiplier from baseline            & 1x - 10x  \\
Antigen tests per day& Multiplier from baseline            & 1x - 10x  \\
Vaccines             & \makecell[l]{Minimum \% of each age group \\\hspace{1em}(5 and older) vaccinated}               & 0 - 75\% of each age group \\
Boosters             & \makecell[l]{Minimum \% of each age group\\\hspace{1em}(5  and older) boosted}  & 0 - 50\% of each age group \\
Contact Tracing      & Number of contacts traced per   day & 6,000 - 60,000             \\
\rowcolor[HTML]{F2F2F2} 
\textit{Behavior}    &   &           \\
Symptomatic Testing OR  & \makecell[l]{Odds ratio to test symptomatic agents\\ \hspace{1em}(relative to non-symptomatic)} & 10 - 100  \\
Testing Quarantine OR& \makecell[l]{Odds ratio to test quarantined agents \\ \hspace{1em}(relative to non-quarantined)} & 1 - 100   \\
Quarantine Adherence Contact   Traced & \makecell[l]{Quarantine adherence when successfully\\\hspace{1em} contact traced} & 0.7 - 1.0 \\
Mask Duration Contact Traced          & When contact traced, wear a mask   (days)            & 0 - 14    \\
Mask Adherence       & Multiplier on mask efficacy         & 0 - 0.2  \\
\bottomrule
\end{tabular}

\caption{Policy descriptions and ranges in simulation experiments. The range of each policy is presented such that the lower end of the range denotes our estimate of the policy intensity at baseline (meant to simulate observed conditions), and the upper end denotes a massive policy increase from baseline conditions. We note that for our vaccination and booster policies, the policies enact a minimum vaccination or booster proportion for each age group, but where empirical data exceeds the minimum, we use the amount given by data. }
    \label{tab:policies}
\end{table}

\subsection{Emulation of Simulation Outcomes}

We simulate two model outcomes using TRACE-STL: the number of cumulative infections, reported at the model's end, as well as the variance in cumulative infections stratified by the Social Vulnerability Index (SVI). Specifically, we calculate the attack rate (number of infections divided by population size) for each SVI category: 0-0.25, 0.25-0.5, 0.5-0.75, and 0.75-1.0, and take the variance. Given a model outcome Y and a dataset X that is n rows x p columns (p=10), our emulation model is a Gaussian process. Specifically, we assume the data generation process:

\begin{align*}
    Y_{i}  \sim N(m_i (X),K_{\theta} (X,X))
\end{align*}

Where $m_i(X)$ denotes a mean function, and $K_{\theta}$ is our covariance matrix, denoting the correlation of our model inputs with one another. In the case of our primary outcome (cumulative infections), since the model outcome is highly nonlinear and varies across a wide range, we first model the mean response with a Gradient-Boosting Machine (GBM), and then fit a heteroskedastic Gaussian process regression to the residual $Y_1- \hat{Y}_1 ^{GBM} \sim N\left(0,K_{\theta}(X,X)\right)$. Thus, for a new input X’, our prediction is:

\begin{align*}
    \hat{Y}_1 = GBM(X')+\mu_{GP} (X')
\end{align*}

Where $\mu_{GP}(X')$ is the mean prediction under a Gaussian process conditional on observed data \cite{binois_practical_2018,ogara_hetgpy_2025,gramacy_surrogates_2020}.

In the case of the secondary outcome (SVI variance) we assume a zero-mean function. For both outcomes we use a Mat\'ern $\nu=5/2$ kernel for the covariance and estimate hyperparameters via maximum likelihood.

\subsection{Generating Variation in Infection Rates by Census Tract}

Throughout the COVID-19 pandemic, the disease burden has fallen unevenly across socio-economic strata, with more socially vulnerable communities being at higher risk of infection \cite{dasgupta_association_2020,kim_social_2020,lopes_combining_2024,mody_quantifying_2022,oates_association_2021}. Contemporary work has postulated that this due at least in part to more vulnerable groups being less able to restrict their mobility (such as remote work, social distancing, and quarantine) \cite{chang_mobility_2021}. Since we do not use individual-level data of infections, we use the following model mechanism to generate heterogeneity in infection risk by SVI. At model instantiation, each agent is given a probability that they will quarantine if they are symptomatic or test positive for COVID-19. This probability is calculated as:
\begin{align*}
    \text{Pr}⁡(\text{quarantine})=0.5*\text{Pr}⁡(\text{quarantine\_adherence})+0.5*(1.0-\text{SVI}_{\text{tract}})
\end{align*}

Where $\text{SVI}_{\text{tract}}$ is the SVI in an agent’s home census tract, meaning that agents in more vulnerable tracts (closer to 1.0) will be less likely or able to quarantine than agents in less vulnerable tracts. Importantly, in the context of policy implementation, this also means that a quarantine-focused policy would have to provide social support and resources to quarantine individuals, as this mechanism explicitly accounts for an agent’s broader social context rather than quarantine compliance simply being about individual choice, a distinction made clear in prior work and is shown again here \cite{hammond_modeling_2021,kerr_controlling_2021}.

\subsection{Sensitivity Analysis}

We analyze the sensitivity of several of our core model parameters to calibration targets, namely: the base transmission rate, presymptomatic proportion, multiplier on initial infections, symptomatic testing odds ratio, and quarantine adherence. These results are available in Figures S1-S3. We also report the induced distribution of secondary infections as a function of our base transmission rate (the mean of which corresponds to $R_0$ in a fully susceptible population with a single index case) in Figure S4.

\subsection{Computational Implementation}

We used TRACE-STL version 2.0.0 for the simulations in this work. TRACE-STL is programmed in Python 3.11. For our emulators, we used hetGPy version 1.0.2 \cite{ogara_hetgpy_2025} for our heteroskedastic Gaussian processes, and LightGBM via optuna version 4.2.0 \cite{akiba_optuna_2019} for our GBM. Our GBM was trained on the mean cumulative infection response across 20 replicates and hyperparameters were selected via 10-fold cross-validation.

\section{Results}
\label{sec:results}
\subsection{Available data is able to generate a status quo model of the Omicron wave}

\begin{figure}
    \centering
    \includegraphics[width=0.9\linewidth]{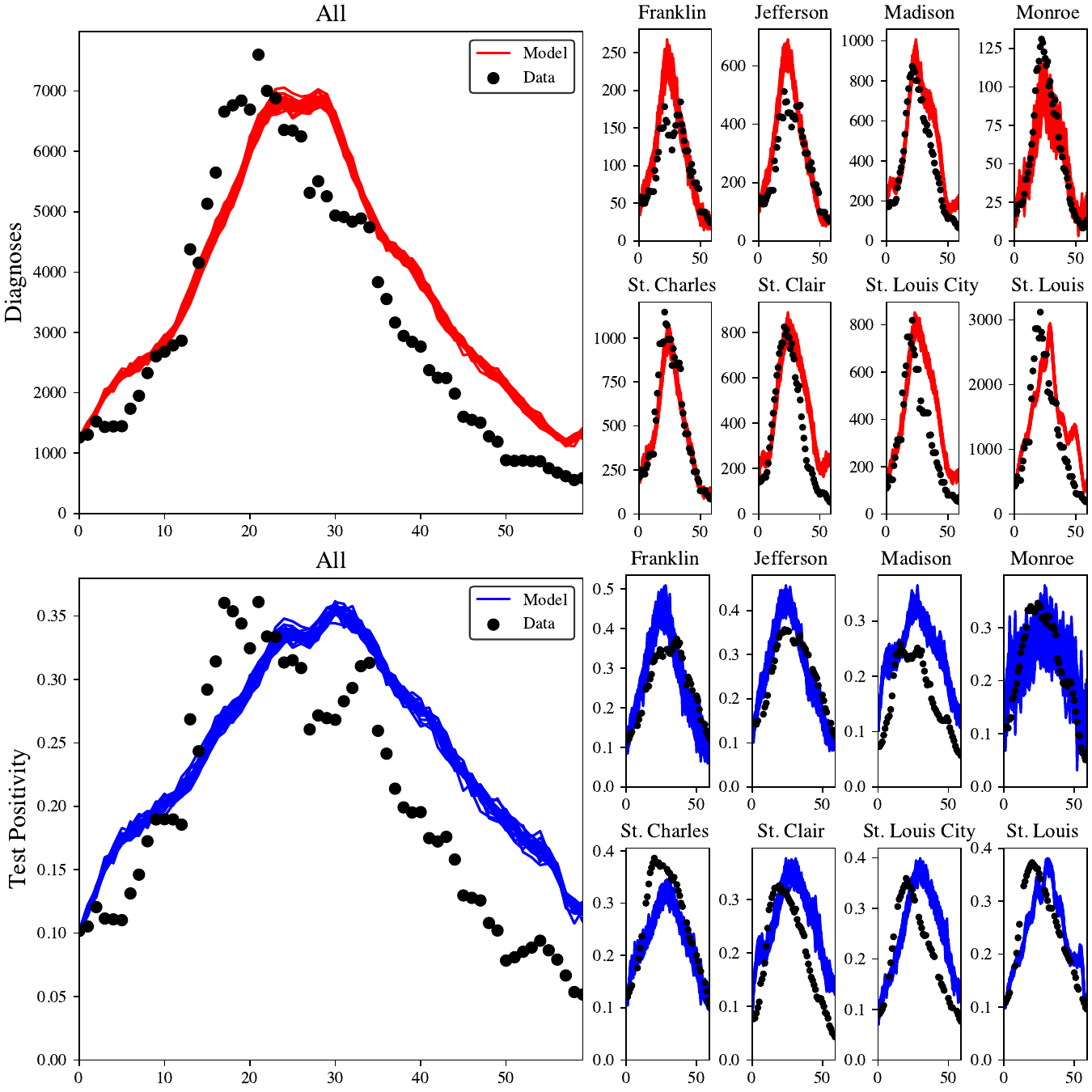}
    \caption{Calibration of model parameters to available data. The top panels show the number of diagnoses per simulated day, at both the population (‘All’) and county levels. Points represent the seven-day average of cases per day, and lines represent model output. The bottom panels show the number of positive tests divided by number of tests reported, at both population and county levels. The model parameters that were varied included the base transmission rate, the number of initially infected agents, and the odds ratio for symptomatic agents to take a test. Their ranges are given in the Supplementary Materials.}
    \label{fig:calib}
\end{figure}

We establish three primary calibration targets for the Omicron wave at two scales of analysis, namely the number of confirmed cases over time, the test positivity rate, and estimated underreporting of cases relative to total infections at the county and aggregate level. In Figure~\ref{fig:calib}, we are able to recover the trend of confirmed cases as well as the test positivity rate across counties and at the population level. We also find that the model reports an average of 842,356 (standard deviation 3,733) cumulative infections and 225,044 confirmed cases (standard deviation 848) compared to 191,759 cases based on CDC data. Our model ratio of 1 in 3.74 cases being reported (confirmed cases divided by infections) is consistent with contemporary estimates during the Omicron wave \cite{institute_for_health_metrics_and_evaluation_ihme_2020,millimet_covid-19_2022,wang_covid-19_2025}. This is driven by the proportion of asymptomatic cases, test availability, and the imperfect likelihood that symptomatic agents will be tested. We find that model stochasticity across populations and random seeds is somewhat low, due to the model’s large number of initially infected agents, as opposed to simulating an outbreak via an index case, which would lead to much higher stochasticity. However, this was intentional in our analysis, because having a large pool of initially infected agents is harder to disrupt, and shifts the focus to mitigation, rather than containment. At baseline, we also find that the mean SVI-stratified cumulative infection rate is 40.53\% (standard deviation 0.27) for the most socially vulnerable census tracts (SVI between 0.75 and 1.0, 106 census tracts with 333,593 agents) to 32.35\% (standard deviation 0.18) for the least socially vulnerable tracts (between 0 and 0.25, 265 census tracts total with 1,043,650 agents).

\subsection{Singular policy interventions would be estimated to reduce disease spread, but would require large increases from baseline}

\begin{figure}[h]
    \centering
    \includegraphics[width=0.9\linewidth]{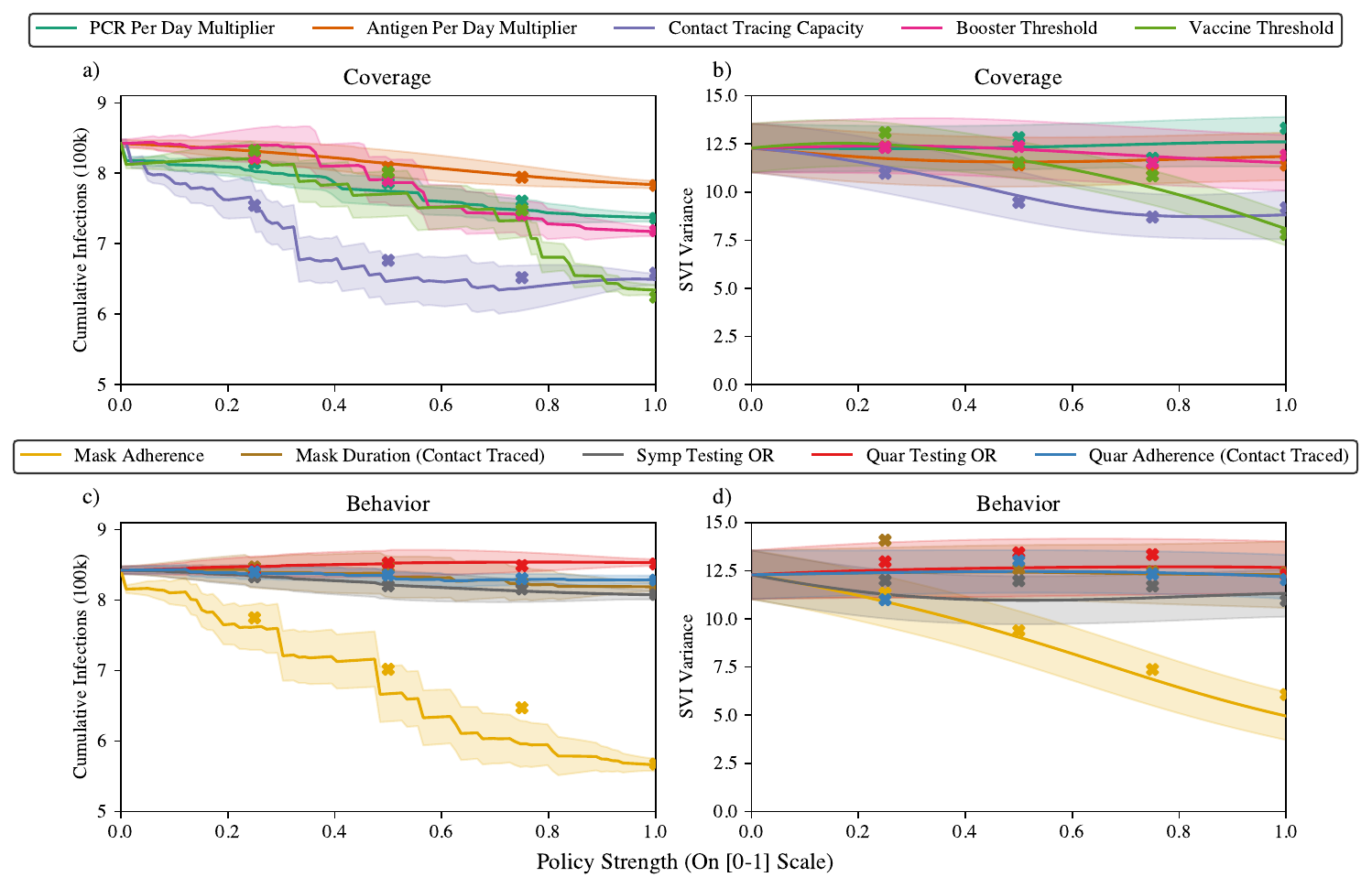}
    \caption{Varying singular policy interventions. Each of our ten policies are varied from their lowest intensity (meant to estimate the policy configuration at baseline) to their highest (representing a substantial increase from baseline) and are estimated via emulator models. Our ten policies are split into two categories in the top and bottom panels: `Coverage' policies are denoted in panels (a) and (b) and `Behavior' policies are denoted in (c) and (d) as discussed in Methods).Y axes represent the cumulative infections over a model run (left panels) and the variance in SVI-stratified disease spread (right panels). For both outcomes, lower values indicate a more favorable outcome (leading to less infection, as well as less variation in disease spread as measured by SVI). Shaded regions represent 90\% predictive intervals, and x's denote validation runs (simulations of policies not explicitly included in the training set with the exception of the points representing the highest policy strength along the righthand edge of each panel). }
    \label{fig:singular-policies}
\end{figure}

Our first experiment explores the efficacy of varying our ten policy interventions over a Latin hypercube sample of size n=1,500 (drawn from a sample of 1,489 locations augmented with baseline conditions, and ten additional parameterizations corresponding to the highest intensity of each policy intervention and all other policies held to baseline) with 20 replicates at each design location for a total of 30,000 simulations. As discussed in Methods, we train a GBM on the mean response of cumulative infections and fit a heteroskedastic Gaussian process regression to the residuals. In Figure~\ref{fig:singular-policies}, we assess the estimated efficacy of each of our emulated policy interventions by varying them to their lowest (status quo) to highest (representing a substantial increase) while holding all other policies at their status quo values. We find that some policy interventions (especially, contact tracing capacity, mask adherence, and vaccine threshold) reduced estimated disease spread (in some cases below 700,000 cumulative infections, and less than 600,000 for mask adherence), but would require large increases from the status quo to accomplish this. We also find that for the emulated interventions shown, a vaccine-focused policy tends to outperform a booster-focused policy if only one can be varied. This comports with a priori expectations since the ceiling for vaccination is higher (a minimum of 75\% of all age groups, as opposed to 50\% for boosters), as well as the potential size of the eligible boosted population being constrained by vaccination in the first place, as well as having a sufficient amount of time since vaccination. For example, we find that setting a target of boosting 50\% of the population but maintaining status quo vaccination numbers results in 39.7\% (standard deviation 0.02) of the population receiving boosters. Finally, we find that none of the policies alone, save for mask adherence, would be expected to reduce the variation in SVI-stratified disease spread. We also generally find our emulated estimates of cumulative infections and SVI-stratified variance are in agreement with simulator outputs, as can be seen in by the validation simulations marked as ``x's'' in the figure, which follow the trend as policy strength increases and mostly fall within the predictive intervals.
\subsection{Combining interventions for interaction effects}

\begin{figure}
    \centering
    \includegraphics[width=0.9\linewidth]{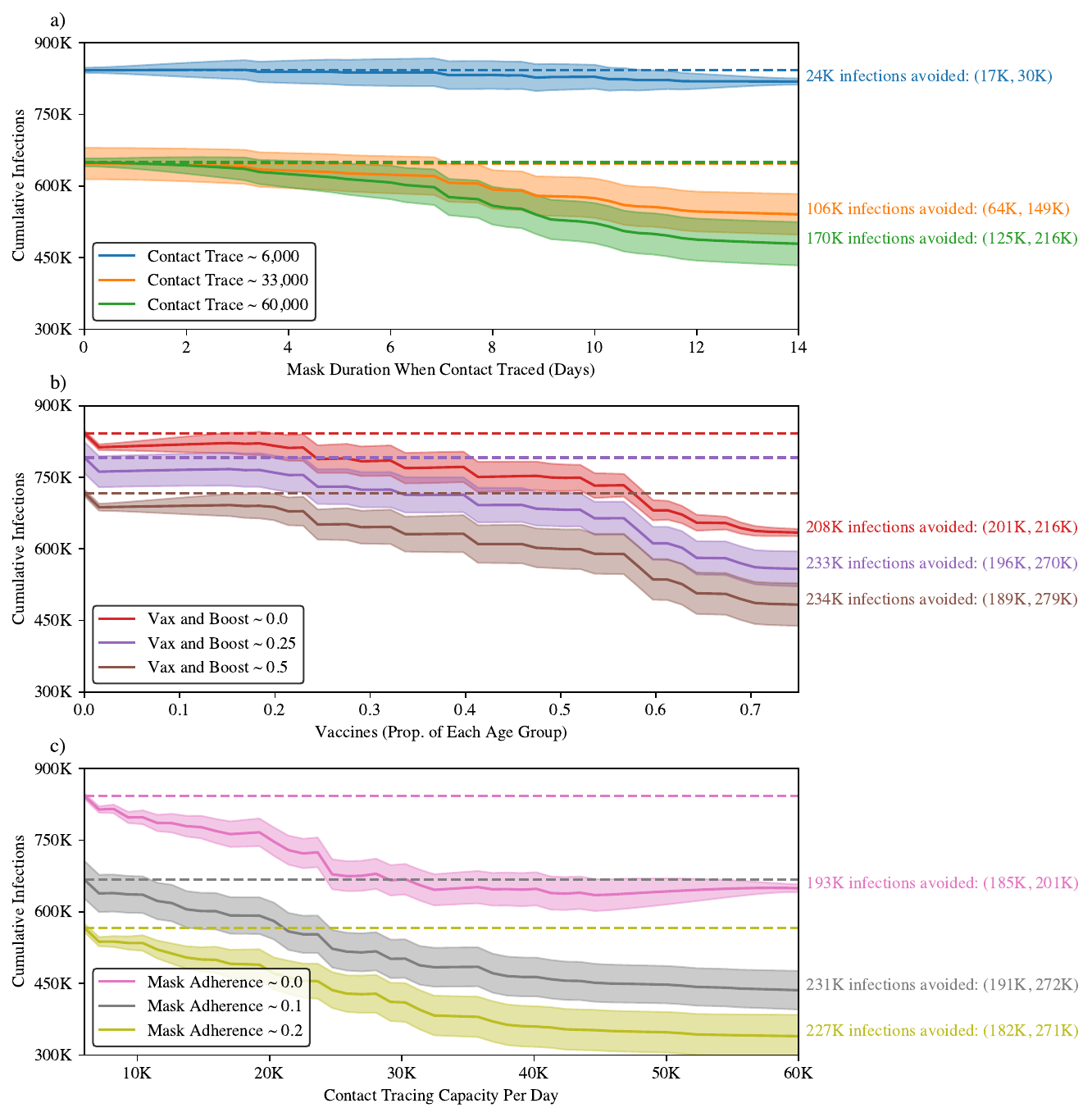}
    \caption{Investigating trade-offs and substitutions when combining policies. In all panels, the Y-axis denotes the cumulative number of infections over a model run (note the truncation at 300K infections). The X-axis in each panel varies a policy from its baseline value through its full policy range. Dashed lines represent the cumulative infections for the affected policy and holding all other policies to baseline. Text on the righthand side of each plot reports the difference in infections holding the X-axis policy at baseline and at its highest intensity for each of the three levels of the policies represented by dashed lines. Shaded regions and text in parentheses represent 90\% predictive intervals. See Table 1 for policy descriptions.}
    \label{fig:interaction}
\end{figure}

The efficacy of several of our policy interventions may be improved when other interventions are present at appropriate intensities. In particular, since our ``time to mask'' has agents wear a mask when they are reached by contact tracing, we anticipate that the with increased contact tracing, we expect the efficacy of that policy to increase. In Figure~\ref{fig:interaction}, we examine three such scenarios. Panel (a) shows the effect of asking agents to wear masks for a certain number of days after being contact traced (from 0 to 14 days) for contact tracing capacities of 6,000, 33,000, and 60,000 agents per day. We can see that the marginal effect increasing ``Time to Mask'' from 0 to 14 days results in an estimated 24,000 infections being avoided at the baseline value of 6,000 for tracing capacity, but 170,000 infections avoided when there is a substantial increase in contact tracing, since the contact tracing process initiates masking in agents who are at a higher risk of being infected, disrupting potential transmission chains. Similarly, panel (b) demonstrates the interplay between initial vaccine availability supporting the efficacy of a booster-focused policy: timely pre-vaccination of at least 75\% of eligible age groups and a booster target of 50\% of those in eligible age groups results in an estimated 234,000 infections being avoided, compared to a booster-focused policy alone, reducing the outbreak size to about 483,000 cumulative infections (an over 42\% decrease from baseline). The emulation process also underscores the opportunities to uncover potential tradeoffs and thresholds, which are shown when vaccination increases from baseline (in red) can substitute for booster doses (in purple) such as in the range of 40\% (where the lines cross). Panel (c) showcases the interaction of the two most effective non-pharmaceutical interventions.

\subsection{Pooling policies demonstrates a large landscape of small policies to mitigate spread}

\begin{figure}
    \centering
    \includegraphics[width=0.9\linewidth]{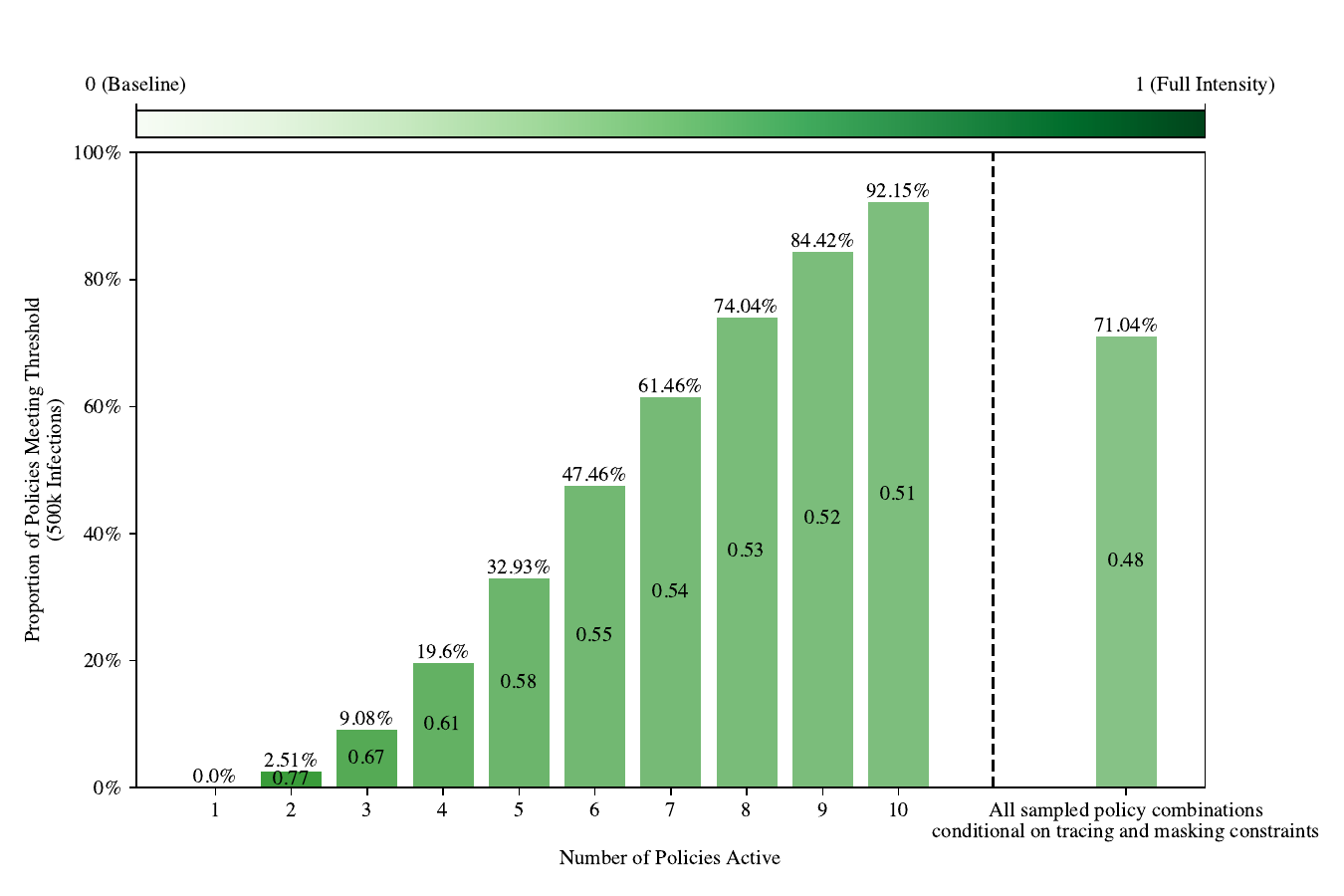}
    \caption{Pooling mixtures of many low-intensity policies estimated to meet policy goals. The x-axis shows the number of policies ``active'' at one time (set to a higher value than baseline), and the y-axis shows the percentage of emulated policies estimated to meet the policy goal of 500,000 cumulative infections or less. Each bar is shaded according to the average policy intensity (0 representing baseline, and 1 representing the largest policy value) of the active policies. For k =10 (in the unconstrained case), we emulate 500,000 potential policies. The rightmost bar represents the scenarios where all ten policies were varied, but only consider candidate interventions with no more than 33,000 contact traced per day and mask adherence no more than 0.1 (half of the range explored for each policy). All other bars contain the results of 5,000*$\binom{10}{k}$ policy emulations where k is the number of active policies. }
    \label{fig:policy-landscape}
\end{figure}

Prior work, both for COVID-19 transmission as well as other communicable diseases has demonstrated that singular policy interventions are often outperformed by combining multiple interventions, especially ones that affect the population in differential ways \cite{germann_mitigation_2006,halloran_modeling_2008,kerr_controlling_2021}. However, the strength of these different interventions must be pre-specified a priori, and when policy options are numerous, modelers are restricted to only testing a few values of each intervention. With our emulation strategy, we can explore far more potential intervention scenarios than would be feasible to simulate with the TRACE-STL model. This allows us to specify model outcomes of interest and search for intervention scenarios that satisfy this criteria. As a motivating example, we investigate the efficacy of the number of interventions deployed at one time, as well as the required intensity of those interventions to meet a policy goal: in this case, no more than 500,000 cumulative infections. These results are shown in Figure~\ref{fig:policy-landscape}, where we sequentially increase the number of ``active'' policies (with intensities higher than their baseline values) and calculate the percentage of estimated policies that meet the policy goal. Specifically, we do the following: For each number of active policies $k \in \left[1,...,9\right]$, we take a Latin Hypercube sample of 5,000 for each combination of $\binom{10}{k}$ policies and fix the remaining policies to baseline. We also calculate the average policy intensity of the active policies. Finally, for $k$=10 we draw a sample of 500,000 candidate policies to emulate a massive landscape of potential intervention strategies.
We see that by increasing the number of policies, we are able to simultaneously identify more candidate policies to meet the policy goal, and that combinations of lower intensity mixtures may be identified. As a sensitivity and robustness check, we also explore the policy combinations that utilize less than 33,000 contacts traced per day and a mask adherence of less than 0.1, each of which were the midpoint of the full range of those policies, since they are individually effective, but increasing them to their full intensity may not be as feasible or cost-effective compared to milder increases of other policies. The percentage of these policies meeting the policy goal is 71.04\%.

\subsection{Policy specification can reduce disease spread while improving geospatial consistency in outcomes}

We validate our policy findings by selecting the 10 smallest policies satisfying both our policy goal and the self-imposed constraint (less than 500,000 infections, no policy using more than the midpoint of contact tracing capacity and mask adherence)and simulating them with TRACE-STL. We select the 10 ``smallest'' policies by ranking them according to the norm of the policy vector:
\begin{align*}
\text{policy combination intensity}= \sum_{i=1}^{10}p_i^2 
\end{align*}

Where $p_i$ represents each individual policy strength on a [0-1] scale, and $p_i=0 \text{ }\forall i$ corresponds to baseline. Note that this differs slightly from the previous analysis where we reported the mean policy intensity: $\frac{1}{k} \sum_{i=1}^k p_i$ where $p_i>0$. This was chosen as a preference for policy combinations that had a mixture of multiple policies (for a two-dimensional example, this is equivalent to saying that combining two policies at half strength would be preferable to doing one at full strength, i.e. that $0.5^2  +0.5^2=0.5\leq0^2+1^2=1$).

\begin{figure}
    \centering
    \includegraphics[width=0.9\linewidth]{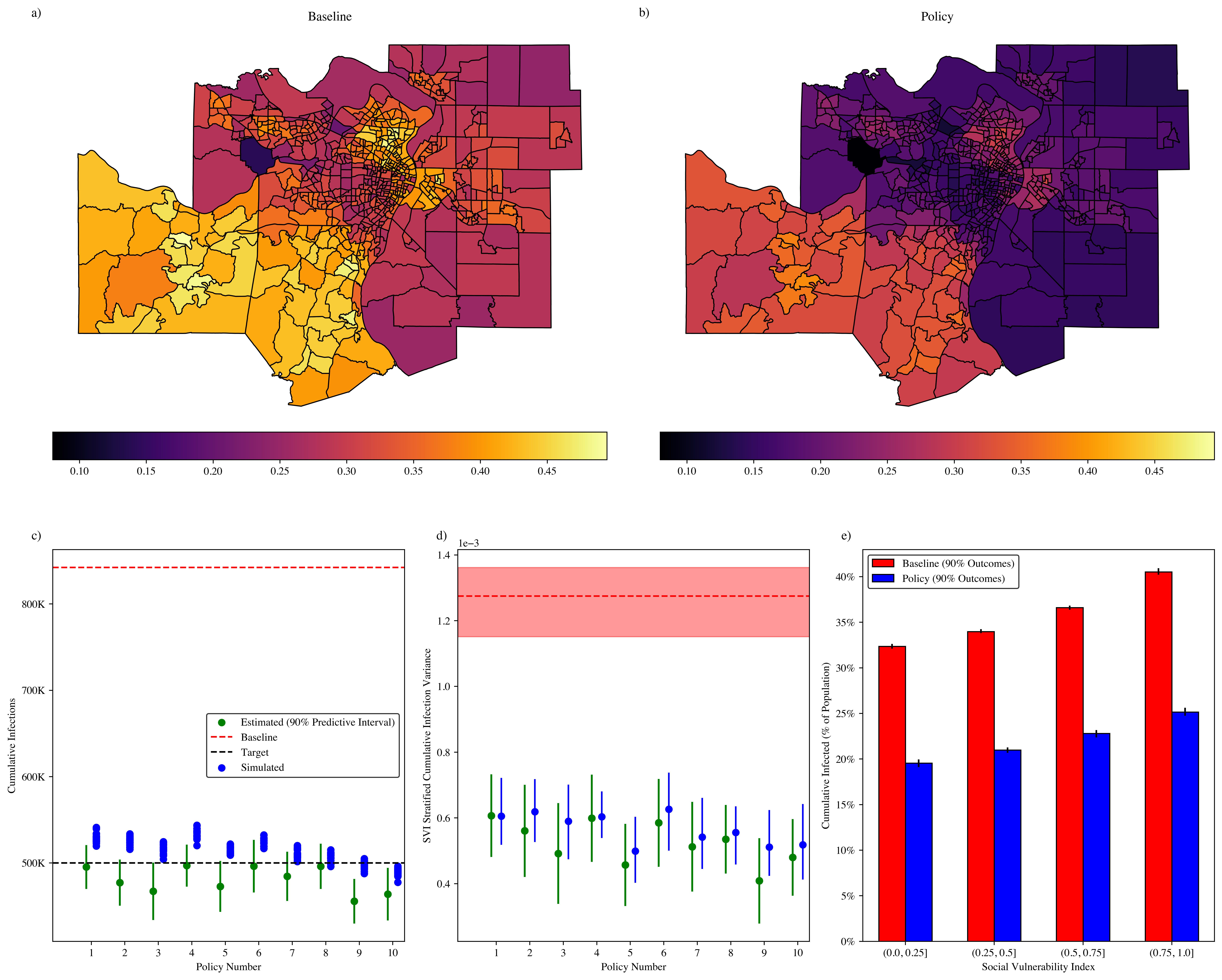}
    \caption{Policy emulation comports with simulation outcomes and can identify geospatially consistent policies. Panel (a) shows census-tract level disease spread at baseline, where color denotes the proportion of agents in each census tract (measured by home location) infected during a model run. Panel (b) shows the same calculation under a policy counterfactual mixing many low-intensity policies together, which achieves substantially lower disease spread as well as lower variance across SVI categories. Our ten simulated policy mixtures are shown in panels (c) and (d), reporting results for both cumulative infection as well as the variance SVI-stratified disease spread. Panel (e) reports the cumulative infection rates by SVI category at baseline and under the policy counterfactual.}
    \label{fig:policy-map}
\end{figure}

In Figure~\ref{fig:policy-map}, we demonstrate that we are able to systematically predict the cumulative infection rate using our emulation strategy compared with those from the model while reducing the variance in SVI-stratified disease spread (panel (c)). Panel (a) shows cumulative disease spread by census tract at baseline, while (b) shows cumulative disease spread under one of our simulated policies (specifically the one corresponding to ``Policy Number 3'' in panels (c) and (d)). The specific policy implementations are available in Table S1, and notably use a relatively high value of contact tracing capacity ($\sim$28,000 contacts per day) supported by a somewhat higher time to mask when contact traced ($\sim$3 days) and higher testing availability ($\sim$2.1x for PCR, $\sim$2.4x for antigen) which allow the contact tracing process to identify more individuals and break the chain of infection for agents who are successfully traced. Since our measure of social vulnerability is based on agent geographic location, we expect a policy that acts on dynamic information, such as contact tracing, to play a role in reducing geospatially-stratified disease spread. Panel (e) emphasizes the difference in SVI-stratified disease spread between baseline and Policy \#3, notably in the reduction for the most socially vulnerable agents. In fact, the variance in SVI-stratified disease spread at baseline (1.27x10$^{-3}$) is more than twice as high as under our policy (5.69x10$^{-4}$). While in some cases our emulated predictions predict a lower total of cumulative infections than are reflected by the simulations, we also note that we intentionally selected the ten ``smallest'' policies estimated to meet the criteria of 500,000 cumulative infections, meaning that these are among the most challenging policies to meet the criteria, since doing more of any policy is expected to reduce the number of cumulative infections.

\section{Discussion}
\label{sec:discuss}
When preparing for disease response, policymakers typically have a short time to act and must choose from several potential intervention scenarios in the face of potentially competing outcome objectives, high complexity, and significant cost. Detailed mechanistic simulation models can help explore interventions and identify the most effective way to intervene given objectives and constraints, but extensive experimentation is often infeasible except for the simplest models. In this work, we have demonstrated the utility of emulator models as a tool to significantly expand the ability of computational models to explore intervention scenarios in complex situations. We show how thoughtful use of emulation can predict simulator outputs at untested input locations, resulting in a more detailed exploration of parameter space. This opens the door to much more detailed exploration of model parameter space, with high potential utility utility for policymakers, since it would allow decisionmakers to ``start from the end'' and specify their outcome objectives ahead of experimentation (such as in our case of filtering to interventions resulting in less than 500,000 infections) and then systematically solve for intervention scenarios that achieve this outcome. 

However, we do note that this ``policy target'' analysis has the benefit of hindsight in the case of the COVID-19 Omicron wave, and choosing an outbreak threshold or outcome before it happens may bring additional challenges in practice.  To combat these difficulties, our emulation framework naturally allows for the exploration of tradeoffs, substitutions, and sensitivity analysis, all of which are critical pieces of information a policymaker would want to have in a high-stakes scenario. In our experiment there were a multitude of low-level policy combinations that met the objective. These findings may be advantageous for policymakers who wish to meet their goals while implementing policies that are tailored to their specific community, especially in cases where community receptiveness to policies varies by context, such as masking and vaccination coverage during the COVID-19 pandemic. Finally, our results here underscore that identifying suitable interventions for complex disease dynamics is still a challenging problem, and even mixtures of many policy interventions that were identified as ``small'' relative to our full search space of model parameterizations still represent somewhat large increases from the status quo observed in data, and underlines the need for policymakers to have substantial resources at their disposal to mitigate disease spread. As mentioned earlier, our focus on identifying the smallest necessary increases from baseline that meet the disease reduction target and reduce variation in geographic disease spread likely have downstream benefits in that they may be less socially disruptive and more cost-effective, and more likely to prevent secondary epidemiological flareups, offering evidence that they may be more efficient than other candidate sets of policy interventions.  

This work could be extended in several ways. We focused on the specific application of a detailed COVID-19 simulation model, but the methods and framework are likely applicable to other types of complex simulation models used for scientific inquiry and to inform policy choices. Of particular interest is further exploration of multi-objective optimization, especially when the objectives are in opposition. Using our epidemiological model as an example, a natural extension would be to model competing objectives such as cumulative disease spread and economic cost, political feasibility, or social disruption (with the model incorporating appropriate assumptions about how these change alongside increases in policy intensity). Recent work has explored the use of emulator models in this context \cite{binois_portfolio_2025}, and extensive exploration of high-dimensional multi-objective complex simulation models represents a promising research direction.

\bibliographystyle{plain}
\bibliography{refs}

\newcommand{\beginsupplement}{
  \newcounter{offset}
  \setcounter{offset}{\value{figure}}
  \newcounter{tbloffset}
  \setcounter{tbloffset}{\value{table}}
  \renewcommand{\thefigure}{S\the\numexpr\value{figure}-\value{offset}\relax}
  \renewcommand{\thetable}
  {S\the\numexpr\value{table}-\value{tbloffset}\relax}
}

\newpage

\beginsupplement

\section{Supplementary Figures}

\begin{figure}[h!]
    \centering
    \includegraphics[width=0.9\linewidth]{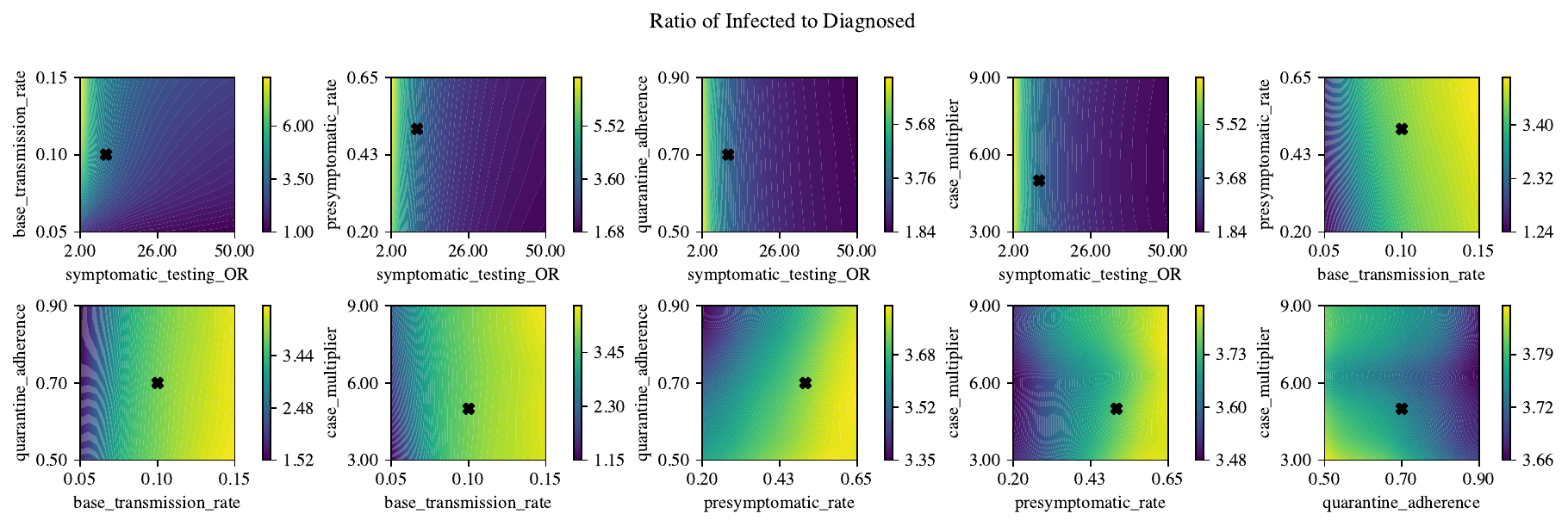}
    \caption{Sensitivity analysis of core model parameters for ratio of infections to diagnoses. Each panel varies two model parameters from their baseline values (marked in ``x'' on each panel). Panels are shaded according to their expected ratio of infections to diagnoses (holding all other model parameters to their baseline values) based on an emulator fit of 500 samples from a Latin Hypercube and 10 model replicates at each parameterization.}
    \label{fig:supp-ratio}
\end{figure}

\begin{figure}
    \centering
    \includegraphics[width=0.9\linewidth]{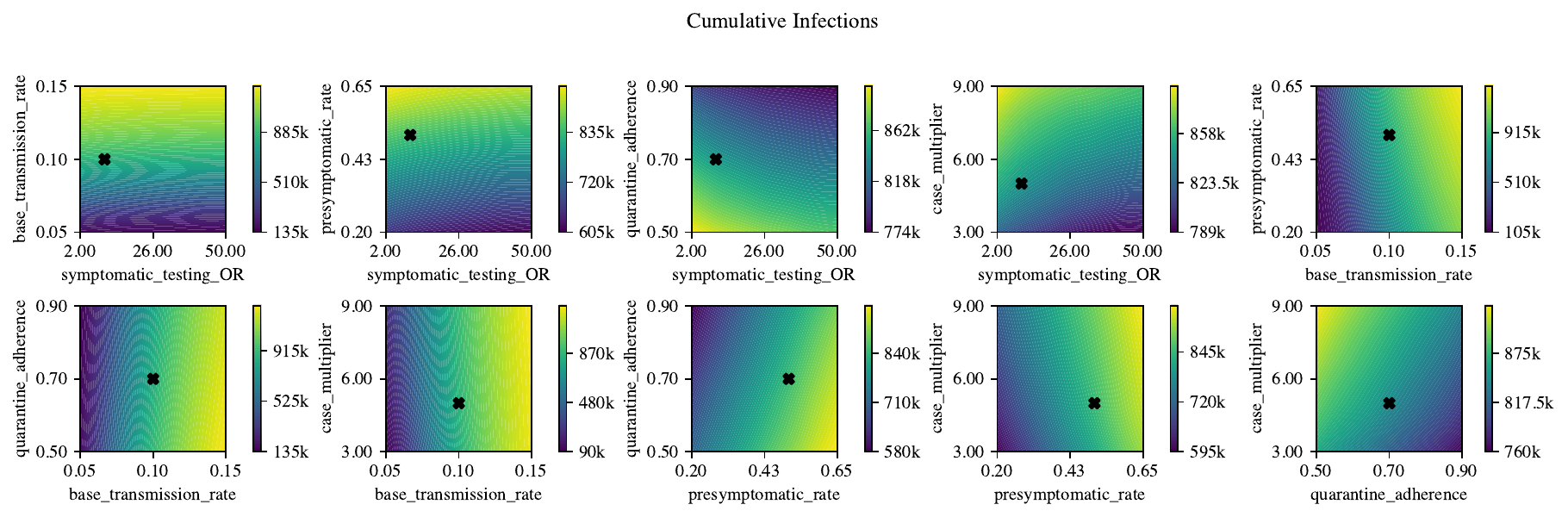}
    \caption{Sensitivity analysis of core model parameters for cumulative infections. Each panel varies two model parameters from their baseline values (marked in ``x'' on each panel). Panels are shaded according to their expected cumulative infections (holding all other model parameters to their baseline values) based on an emulator fit of 500 samples from a Latin Hypercube and 10 model replicates at each parameterization.}
    \label{fig:supp-infections}
\end{figure}

\begin{figure}
    \centering
    \includegraphics[width=0.9\linewidth]{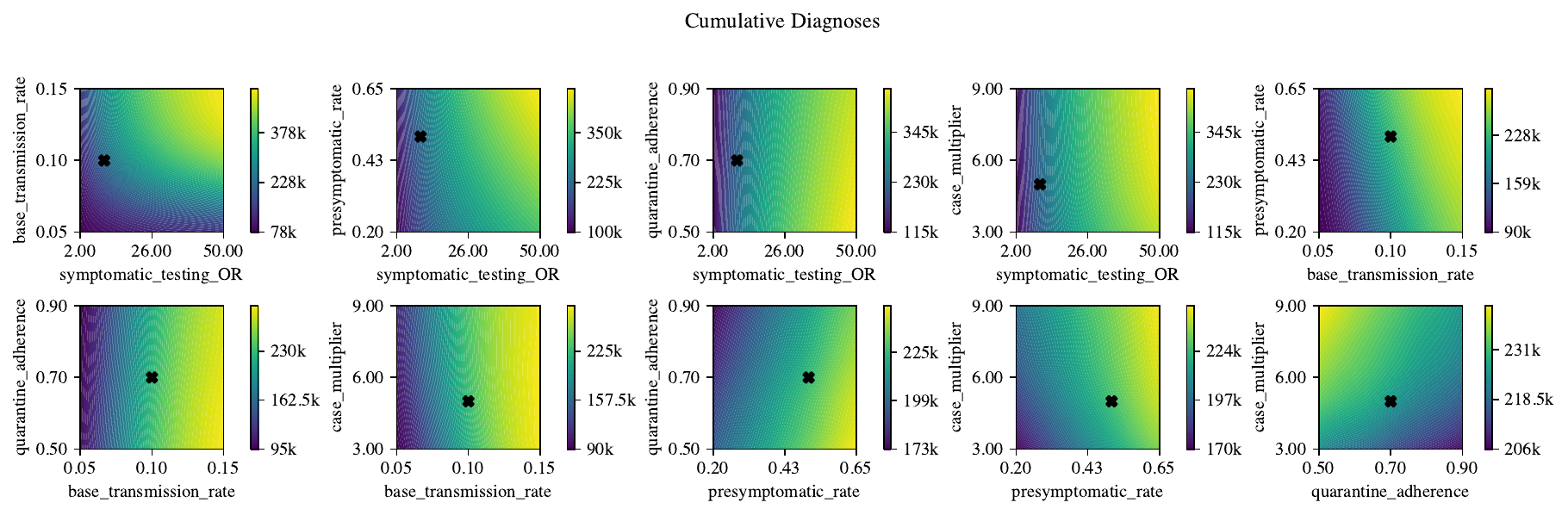}
    \caption{Sensitivity analysis of core model parameters for cumulative diagnoses. Each panel varies two model parameters from their baseline values (marked in ``x'' on each panel). Panels are shaded according to their expected cumulative diagnoses (holding all other model parameters to their baseline values) based on an emulator fit of 500 samples from a Latin Hypercube and 10 model replicates at each parameterization.}
    \label{fig:supp-diagnoses}
\end{figure}

\begin{figure}
    \centering
    \includegraphics[width=0.9\linewidth]{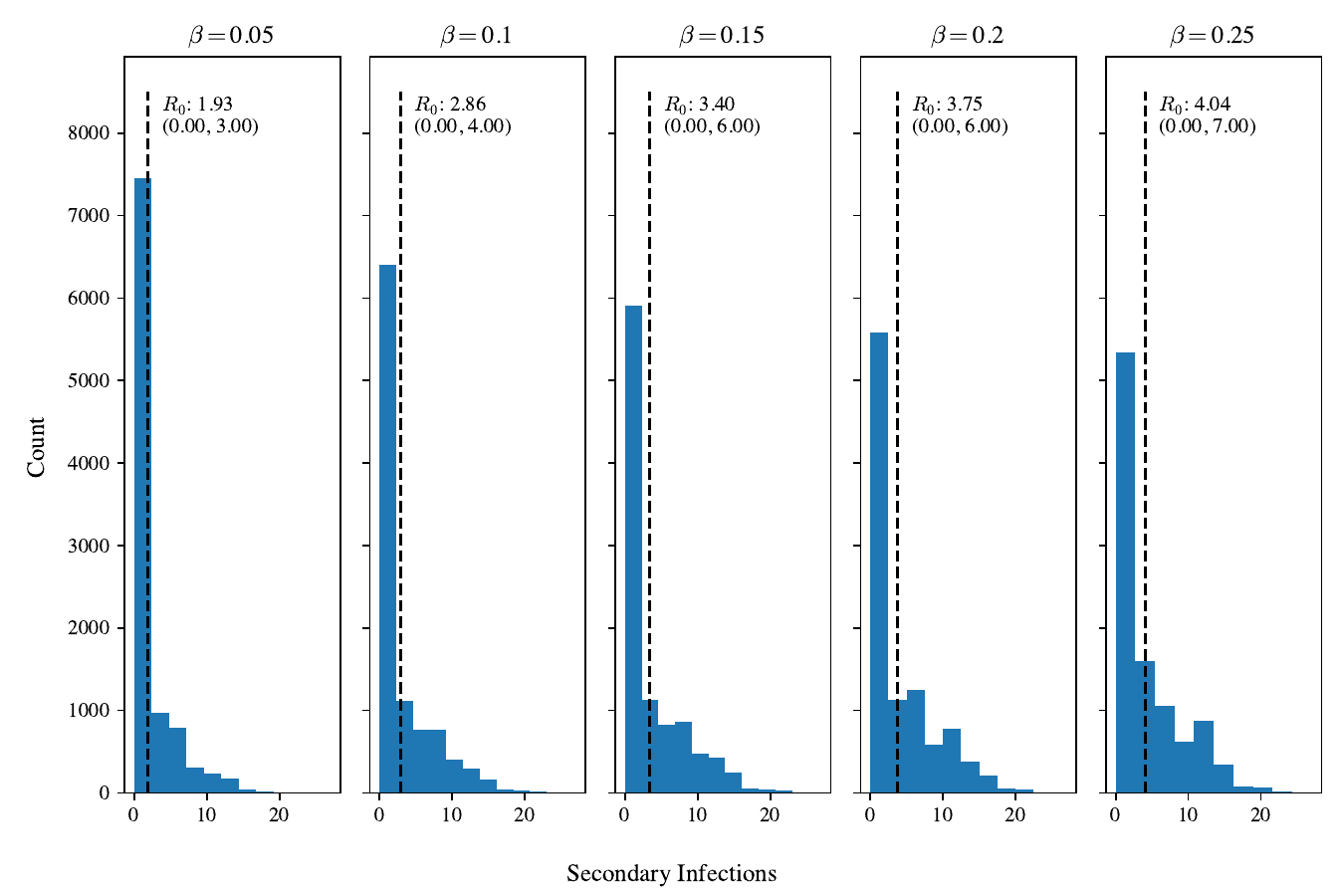}
    \caption{$R_0$ calibration for base transmission rate. Each panel shows a value for the average base transmission rate and 10,000 model simulations of an index case in an otherwise susceptible population. The x-axis reports the number of secondary infections induced by the base transmission rate. Panel labels show the average number of secondary infections, and the middle 50\% (25th and 75th percentile) of secondary infections in parentheses.}
    \label{fig:supp-R0}
\end{figure}

\begin{table}[]
\begin{tabular}[\textwidth]{lllllllllll}
Index & \begin{tabular}[c]{@{}l@{}}PCR\\ Mult\end{tabular} & \begin{tabular}[c]{@{}l@{}}Antigen\\ Mult\end{tabular} & \begin{tabular}[c]{@{}l@{}}CT\\ Capacity\end{tabular} & \begin{tabular}[c]{@{}l@{}}Booster\\ Threshold\end{tabular} & \begin{tabular}[c]{@{}l@{}}Vaccine\\ Threshold\end{tabular} & \begin{tabular}[c]{@{}l@{}}Mask\\ Adherence\end{tabular} & \begin{tabular}[c]{@{}l@{}}Mask\\Duration\\ (CT)\end{tabular} & \begin{tabular}[c]{@{}l@{}}Symp\\OR\end{tabular} & \begin{tabular}[c]{@{}l@{}}Quar\\ OR\end{tabular} & \begin{tabular}[c]{@{}l@{}}Quar\\ Adherence\\ (CT)\end{tabular} \\
\midrule \\
\rowcolor[HTML]{F2F2F2} 
1  & 2.01 & 2.86 & 27,854& 0.1& 0.27  & 0.08  & 2  & 15.56  & 16.21  & 0.7 \\
2  & 1.78 & 2.92 & 30,809& 0.14  & 0.05  & 0.07  & 4  & 27.9& 7.4 & 0.73\\
\rowcolor[HTML]{F2F2F2} 
3  & 2.16 & 1.45 & 22,641& 0.25  & 0  & 0.09  & 2  & 12.8& 5.81& 0.78\\
4  & 1.15 & 1.79 & 28,274& 0.07  & 0.07  & 0.09  & 3  & 26.55  & 14.25  & 0.81\\
\rowcolor[HTML]{F2F2F2} 
5  & 2.66 & 4.14 & 28,294& 0.02  & 0.21  & 0.06  & 3  & 29.39  & 19.71  & 0.76\\
6  & 1.39 & 4.62 & 23,869& 0.18  & 0.16  & 0.07  & 3  & 32.69  & 6.87& 0.75\\
\rowcolor[HTML]{F2F2F2} 
7  & 3.88 & 1.31 & 26,112& 0.1& 0.03  & 0.07  & 4  & 36.11  & 8.23& 0.8 \\
8  & 1.33 & 1.25 & 29,677& 0.16  & 0.02  & 0.1& 0  & 12.54  & 28.95  & 0.82\\
\rowcolor[HTML]{F2F2F2} 
9  & 3.72 & 1.96 & 32,096& 0.07  & 0.02  & 0.09  & 0  & 17.05  & 21.45  & 0.81\\
10 & 1.18 & 1.69 & 29,974& 0.17  & 0.13  & 0.1& 3  & 32.62  & 24.5& 0.74\\
\rowcolor[HTML]{F2F2F2} 
Range & {[}1,10{]} & {[}1,10{]} & {[}6k,60k{]}& {[}0.0,0.5{]}  & {[}0.0,0.75{]} & {[}0,0.2{]} & {[}0,14{]}  & {[}10,100{]} & {[}1,100{]}  & {[}0.7,0.9{]}  \\
\bottomrule
\end{tabular}
\caption{Policy values for geospatially consistent policies. Table values show the specific policy parameterizations for the 10 “lowest intensity” policy combinations estimated to meet the policy goal of 500,000 cumulative infections or less. Policy descriptions are available in Table 1 in the main text.}
\end{table}

\clearpage



\section*{Supplementary material}

# TRACE-STL MODEL DESCRIPTION

VERSION 2.0

**David O'Gara et. al**

## 1 Introduction

The TRACE framework is a series of agent-based models that have informed policy response planning during and after the COVID-19 pandemic [16, 32]. This document summarizes TRACE-STL version 2.0, a model built for the metro area in St. Louis, Missouri. Previous technical appendices are available in the previously-mentioned published works, and this document may be seen as a continuation of those.

### 1.1 Initialization

Agents in the model are each initialized with a disease state ($S$ = susceptible, $E$ = exposed, $P$ = presymptomatic, $I$ = symptomatic, $A$ = asymptomatic, $R$ = recovered). The state transitions are the same as Figure 1 of [32], and is reproduced here as Figure 1:

#### 1.1.1 Initial State Estimation

TRACE-STL initializes agents to be representative of the state of COVID-19 in the St. Louis area. The number of agents in each disease state is calculated based on case data for the relevant period and adjusted for an initial under-reporting using a case multiplier (in our experiments, 5.0 and varied in sensitivity analysis). Specifically, we:

- Calculate the number of infections over the last $I$ days before model instantiation, where $I$ is the average infectious duration
- Aportion the agents into infection states where they are Presymptomatic, Asymptomatic, or Symptomatic
- Assign each infected agent a "time since infection" (uniformly between 0 and the infection state transition duration)
- Adjust the number of "new diagnoses" (confirmed cases via positive diagnostic test) in the model on day $t = 0$ to match empirical data.

During model initialization, we allow agents who have been previously infected before the Omicron wave to probabilistically move from Recovered back to Susceptible (see Infection State Transitions for more detail). This results in about 75% of the agent population being Susceptible to infection upon model instantiation, though this does depend on model parameters.

#### 1.1.2 Contacts

The home, school, workplace, and community contact structures are the same as the ones used in TRACE-STL version 1.0 [16]. For St. Louis City and St. Louis County, which reflect more urban areas relative to the remaining counties, we add a small number of additional 'community' contacts. The number of additional contacts per agent is calculated with a random draw from a Poisson distribution with rate parameter of 0.5.

### 1.2 Infection Dynamics

Each simulated day $t$, each agent $i$ performs the following actions:

1. Interact with all agents $j$ on their current active contact list ($C_{it}$)

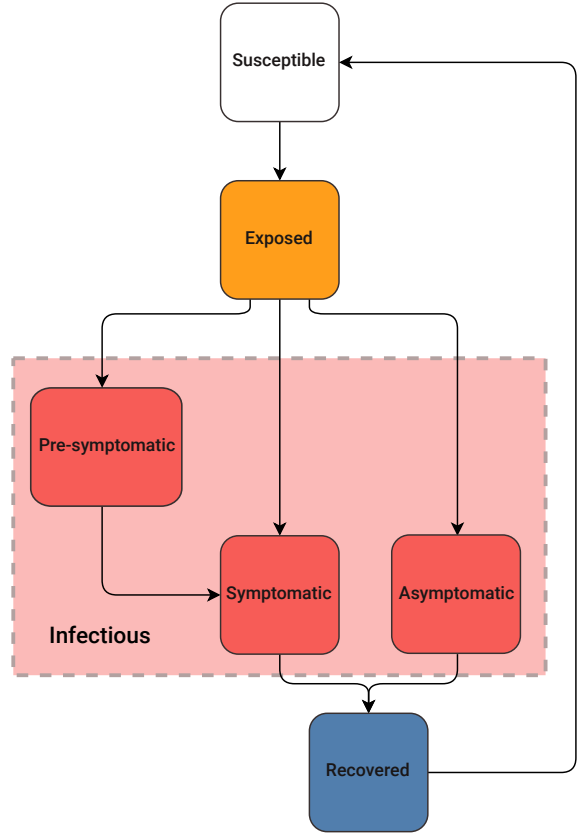

Figure 1: Summary of agent state transition pathways. Figure used with permission from [32].

2. If $i$ is susceptible , for each interaction with an infectious (state $A$, $P$, or $I$) contact $j$ in a setting that $j$ currently is not removed from due to social distancing, $i$ has a probability of being infected by $j$.

This calculation is as follows:

$$Pr(\text{infection}_{ijt}) = T_j * A_j A^{'} * M_j M^{'}_{out} * M_i M^{'}_{in} * V_{in}(w_i, v_i) * V_{out}(v_j)$$

Where:

- $T_j$ is the transmission value for agent $j$. We allow for heterogeneous infectivity (i.e. ability to transmit disease during contact) for agents in an infectious state [11].
- $A_j$ indicates whether $j$ is asymptomatic, and $A'$ is an asymptomatic transmission rate multiplier
- $M_j$ indicates whether $j$ is wearing a mask and $M^{'}_{out}$ is the multiplier for efficacy of reducing infectivity for a mask-wearing potential infector.
- $M_i$ indicates whether $i$ is wearing a mask and $M_{in}$ is the protective multiplier for of reducing infection probability for a mask-wearing potential infectee.
- $V_{in}(w_i, v_i)$ is a function that outputs a multiplier given the vaccine or booster type that $i$ has, $v_i$, and the number of days since $i$ completed their initial vaccine series or since they were last boosted, $w_i$. If $i$ is unvaccinated this multiplier is 1.
- $V_j(v_j)$ is a function that outputs the correct multiplier for the given vaccine that $j$ has $v_j$.

### 1.2.1 Infection State Transitions

1. When an exposure occurs for susceptible agent $i$ occurs they move into state $E$

2. At time of exposure they are pre-designated an infectious state:
   (a) $A$ with probability $A^*$ (the asymptomatic rate)
   (b) $P$ with probability $P^*$ (the presymptomatic rate)
   (c) $I$ with probability $1 - A^* - P^*$
3. After the latent duration appropriate for their predesignated infectious state, agents transition to that state ($A$, $P$, or $I$). In version 2.0, the latent duration is assigned for each agent with a random draw from a gamma distribution with shape and scale $\Gamma(4.09, 0.875)$ [13].
4. If an agent is in state $P$, they transition to state $I$ after the period of pre-symptomatic duration.
5. If agent i is in state $A$ or $I$, it will transition $R$ to state after the disease duration has passed. In version 2.0, an agent's infectious duration is assigned with a random draw from a gamma distribution with shape and scale $\Gamma(6.62, 0.83)$ [17].
6. After a certain time (in our experiments, 90 days), recovered agents have a probability each day of transitioning back to susceptible based on draws from a geometric distribution with $p = \frac{1}{480}$
90 days represents the time to peak antibody response after infection and 480 days represents the median time to reinfection (about 16 months) [39].

## 1.3 Parameterization

The full list of constant parameters in TRACE-Omicron is described below in Table 1

| Parameter | Value | Source & Notes |
|---|---:|---|
| *Epidemiology* | | |
| Base Transmission Rate | 0.10 | Calibrated |
| Asymptomatic Rate | 0.33 | [25] |
| Presymptomatic Rate | 0.5 | [9, 23, 17]* |
| Asymptomatic Factor | 1.0 | [23] |
| Latent Duration of Initial Cases (Days) | 3 | [20] |
| Presymptomatic Duration of Initial Cases (Days) | 2 | [37, 14] |
| Infectious Duration of Initial Cases (Days) | 5 | [11, 31] |
| Initial Case Multiplier | 5 | Calibrated |
| Historic case multiplier | 4.5 | [7, 19] |
| Time to Lost Immunity After Peak Immunity (Days) | 480 | [39] |
| Time to Peak Immunity (Days) | 90 | [39] |
| *Quarantine* | | |
| Quarantine Adherence | 0.7 | [1, 10, 40] |
| Quarantine Length | 10 | [6] |
| Quarantine Length (Asymptomatics) | 5 | [6] |
| *Testing* | | |
| PCR False Negative Rate | 0.03 | [36] |
| PCR False Positive Rate | 0.05 | [38] |
| Antigen False Negative Rate | 0.4 | [29, 22] |
| Antigen False Positive Rate | 0.05 | [29, 22, 5] |
| Antigen Reporting Multiplier | 2.5 | [34] |
| Contact Testing Odds Ratio | 1.0 | Assumption |
| PCR Reporting Delay (Days) | 0 | Assumption |
| Antigen Reporting Delay (Days) | 0 | Assumption |
| *Masking* | | |
| Masking Infection Protection | 0.8 | [41, 4, 8] |
| Masking Transmission Protection | 0.7 | [41, 4, 8] |
| *Vaccination* | | |
| Pfizer, Moderna, J&J Vaccine Proportions | 0.6, 0.3, 0.1 | [28] |
| Pfizer, Moderna, J&J Vaccine Efficacy | 0.505, 0.488, 0.532 | [26] |
| Pfizer, Moderna, J&J Booster Efficacy | 0.6475, 0.495, 0.6475 | [26] |
| Pfizer, Moderna, J&J Vaccine Waning Rates | 0.0071, 0.0053, 0.004 | [26] |
| Pfizer, Moderna, J&J Booster Waning Rates | 0.004135, 0.00283, 0.004135 | [26] |
| Pfizer, Moderna, J&J Vaccine Efficacy Against Transmission | 0.25, 0.25, 0.15 | [26] |
| Pfizer, Moderna, J&J Booster Efficacy Against Transmission | 0.5, 0.5, 0.5 | [3] |

*Note: we assume that half of all symptomatic cases may transmit during the presymptomatic stage. Widely used models such as [23, 17] assume that most or all symptomatic agents may transmit in the presymptomatic stage. Our formulation is thus a compromise between our previous work [16] which assumed a presymptomatic rate of 20%.

Table 1: Constant parameters in TRACE-STL. These parameters may be combined with Table 1 in the main text for a full characterization of the model parameters.

## 2 Modeling COVID-19 Containment Policies and Practices

### 2.1 Social Distancing

We represent social distancing measures in our model as:

1. **Quarantine:** agents who believe that they are infected, receive a positive COVID-19 test result (described below), or are identified as being in contact with someone who has been infected (described below) are asked to self-isolate for a period of time determined by the quarantine duration parameter and whether they are symptomatic; they comply with this request probabilistically based on the quarantine adherence parameter and the Social Vulnerability Index of their home census tract. When doing so, they do not have contact with agents other than their household contacts. If non-adherent, they interact with other agents as normal.
2. **School social distancing:** agents remove a sample of their school contacts.
3. **Remote work:** agents selected for remote work do not have workplace contacts.
4. **Community distancing:** agents remove a sample of their contacts in the "other" setting.

Remote work and community distancing are all calculated using county-level Google mobility reports [15]. For school social distancing, we assume a 10% reduction in contacts due to schools being open during the study period, but likely had social distancing and masking protocols in place[30].

### 2.2 Testing and Tracing

Testing and tracing takes place as follows:

1. A sample agents – up to the number of available daily tests per county are tested based on their testing probability (described for each testing strategy below). A test returns a positive result with some probability of false negative results for infected agents, and false positive results for non-infected agents.
2. Agents who test positive are asked to quarantine.
3. Agents who test positive report the members of their daily contact list. A full list of contacts to trace is created from all such contact lists, and agents on this list up to the available daily tracing capacity are asked to self-isolate. Any remaining contacts remain untraced, but may be traced on subsequent days.

Our PCR testing strategy is:

- Each agent is assigned a testing probability (uniform among all agents). Agents are then assigned odds ratios which reflect their likelihood of being tested whether they are symptomatic, contacts of symptomatic, or quarantined.

Antigen testing implements a parallel but mechanically similar testing function to PCR testing. All agents that test positive from either testing mechanism go into the same queue for contact tracing.

Our antigen testing strategy is identical to the PCR testing strategy, with two exceptions:

- Agents do not antigen test when quarantined
- Agents may only test positive if they are symptomatic
- Some antigen tests go unreported. We assume that 60% of antigen tests go unreported [34].

To calculate the number of tests per county per day, we use data from the Missouri Department of Health and Senior Services [27] and the Illinois Department of Public Health [18]. For Illinois testing data, we assume that half are PCR tests and half are antigen. To reflect the assumption that 60% of antigen tests are unreported (see above), we apply a multiplier of $\frac{1}{0.4} = 2.5$ to the amount of antigen tests. We calculate the positivity rate using data from the github repository maintained by Johns Hopkins University [21].

### 2.3 Masking

Agents can wear masks, which reduces their probability of infection relative to an unvaccinated agent, and reduces the probability of their contact being infected based on mask efficacy. Mask multipliers are only applied to non-home contact interactions. Agents may also wear masks for a fixed amount of days at a very high mask adherence (set to be even more effective than N95 masks) [41] when successfully contact traced.

To calculate average masking rates, we use county-level masking data from the `covidcast` library [35, 2]

To model mask efficacy, we model the following:

- Baseline infection and transmission reductions: represents the estimated risk reduction in infection and transmission for cloth masks. Estimates of risk reduction were taken from [41, 4] though we note there is substantial heterogeneity in the literature for the efficacy of masking in observational settings [8]. Further, we also note that our parameters of mask efficacy may be biased upwards (less effective) since we assume that agents who wear masks wear them outside the home at all times (for computational efficiency).
- Mask adherence reflects the usage of higher protection masks (such as N95 masks) as well as well-fitting masks, modeled as a multiplier on mask efficacy. At baseline, this value is 0 (no change), and varies to a 20% additional reduction in risk.
- Therefore, when considering the various mask efficacy multipliers, the full 'effect' of masking varies between a 30% reduction (a susceptible agent wearing a mask interacting with an unmasked infectious agent) to 65% (both infector and infectee wearing a mask at max efficacy), which is in the range of contemporary estimates [8, 41, 24] and estimates used in other modeling work [33].

## 2.4 Vaccination and Boosting

Agents can be vaccinated: when this occurs, they receive a vaccine efficacy multiplier which reduces their probability of infection relative to an unvaccinated agent. Vaccination also reduces the infectivity of agents who are infected despite vaccination. Vaccines are distributed proportionally by age group based on CDC data [12]. An agent's vaccine efficacy multiplier decreases over time depending on the type of vaccine they received (Pfizer, Moderna, or Johnson & Johnson), and how long ago they were vaccinated [26] subject to the model $A \exp(-wt)$ where $A$ is the maximum vaccine protection (achieved after 14 days), $w$ is a waning parameter, and $t$ is the time since vaccination.

Agents can be boosted after 180 days, if their age group is eligible. When they are boosted, their vaccine efficacy multiplier is restored to its original value.

Policy analyses explore preemptive vaccination and boosting, governed by a threshold parameters between 0 and 1, where we vaccinate or boost at least that proportion of each age group. If CDC data reflects a higher initial proportion of agents than the vaccine or booster threshold, then we use the estimate from data. In our analyses, we do not include 0-5 year olds for preemptive vaccination and boosting.



\end{document}